# High Field Magneto-Transport of Mixed Topological Insulators $Bi_2Se_{3-x}Te_x$ (x = 0, 1, 2 & 3)

Deepak Sharma[1,2], Yogesh Kumar[1,2], P. Kumar[3], Vipin Nagpal[3], S. Patnaik[3], V.P.S. Awana[1,2, *]

[1]*National Physical Laboratory, Dr. K.S. Krishnan Marg, New Delhi-110012, India*
[2]*Academy of Scientific and Innovative Research (AcSIR), Ghaziabad, 201002, India*
[3]*School of Physical Sciences, Jawaharlal Nehru University, New Delhi-110067, India*

**Abstract**

The article comprises structural, microstructural, and physical properties analysis of $Bi_2Se_{3-x}Te_x$ (x= 0, 1, 2 & 3) mixed topological insulator (MTI) single crystals. All the crystals were grown through a well-optimized solid-state reaction route via the self-flux method. These MTI are well characterized through XRD (X-ray Diffraction), SEM (Scanning Electron Microscopy), EDAX (Energy Dispersive spectroscopy), and thereby, the physical properties are analyzed through the RT (Resistance vs temperature) down to 10K as well as the magneto-resistance (MR) measurements (at 5K) in a magnetic field of up to ± 10 Tesla. The MR% drastically varies from x=0 to x=3 in MTI, from a huge 400%, it goes down to 20% and 5% and eventually back to 315%. This fascinated behaviour of MR is explained in this article through HLN (Hikami-Larkin-Nagaoka) equation and an additional term. This article not only proposed the mesmerizing behaviour of MR in MTI but also explains the reason through competing WAL (Weak Anti-Localization) and WL (Weak Localization) conduction processes.

[*]**Corresponding Author**

Dr. V. P. S. Awana:  E-mail: awana@nplindia.org
Ph. +91-11-45609357, Fax-+91-11-45609310
Homepage: awanavps.webs.com

## Introduction

Quantum materials such as Topological Insulators (TIs) prosperous the field of spintronic and quantum computation [1–3]. Its surface states (SS) are topologically protected with π Berry phase and supports the dissipationless flow of current due to the presence of band inversion in materials possessing strong spin-orbit coupling (SOC) [4–9] while its bulk behaves as an insulating system [10–12]. These SS are protected by the time-reversal symmetry (TRS) [12] and the SOC in TI is the reason for such kinds of states in the system. The states are robust against imperfections and non-magnetic impurities. The topologically protected SS are present between the bulk conduction band (BCB), and bulk valance band (BVB). These topological properties have been experimentally confirmed through the presence of Dirac cone in various TIs such as $Bi_2Se_3$ and $Bi_2Te_3$ using angle-resolved photoelectron spectroscopy (ARPES). $Bi_2Se_3$ and $Bi_2Te_3$ are the 3D TI which are binary tetradymite compounds [4,9,11]. They exhibit strong SOC and their respective high MR helps to fabricate more reliable magnetic memory based devices, magnetic sensors, to observe quantum oscillations such as Shubnikov-de Hass oscillations [13–15] and some rare oscillations like log periodic quantum oscillations [16]. Recently, non-saturating liner magneto-resistance (LMR) has been observed in monolayer graphene [17], $Bi_2Se_2Te$ and $BiSbTeSe_{1.6}$ thin films [18]. In $Ru_2Sn_3$ single crystals the behaviour of MR changes from quadratic to linear with decreasing temperature [19] and parabolic MR is shown in Cu- and Cr- doped $Bi_2Se_3$ films [20]. Apart from this type of conventional TIs, a new class has also emerged from it, which is considered as mix topological insulators (MTI) [21–24]. Magneto-resistance in MTIs is quite different from that in pure TIs as these do not exhibit cusp like positive MR [25] at low magnetic field. The two basic examples of MTI are $Bi_2Se_2Te$ and $Bi_2Se_1Te_2$. They are ternary tetradymite compounds in which a single Dirac cone exists with a single Dirac point on their helical topologically protected SS [21–24]. The topological behaviour of these MTI can be seen through various physical measurements which include RT, MR measurements, etc. Basically, the experimental results of MR provide vast knowledge about the TI with the help of theoretical fitting through the HLN equation. It includes how their properties are bludgeoned through the surface and bulk contribution as a function of the magnetic field. The HLN is effective in studying the signature of WAL in 2D systems. The HLN parameters viz., pre-factor($\alpha$), and $L_\varphi$ give us information about the type of localization and number of conducting carrier channels in the material. The magneto-conductivity (MC) of TIs can be understood in terms of competing for WAL and WL effects.

In this article, the $Bi_2Se_{3-x}Te_x$ (x= 0, 1, 2 & 3) mixed topological insulator (MTI) single crystals have been synthesized and well-characterized through XRD, SEM and EDAX. The RT measurements have been carried out from room temperature to down to 5K. The MR has been performed at this low temperature of 5K which is in a magnetic field of range up to ± 10 Tesla. The magneto-conductivity data have been analyzed by fitting the Hikami-Larkin-Nagaoka (HLN) equation. HLN curve shows a superlative fit for magneto-conductivity data at a lower magnetic field in the case of pure TI crystals, whereas, for MTI, it showed a large deviation from experimental data. Therefore, we fitted the mixed TIs with a quantum diffusive term which is a combination of SOC and elastic scattering [26]. This indicates the dominating bulk contribution towards MR in mixed TIs over surface one.

**Experimental Details**

Bismuth (Bi) based TI and MTI single crystals have synthesized by using the self-flux method via solid-state reaction route [7,8]. The optimal heat treatment for the MTI have been discussed in our previous reports[27]. The XRD measurements have been performed on mechanically cleaved single-crystal flakes using the Rigaku X-ray diffractometer with the $CuK_\alpha$ radiation ($\lambda$ = 1.5418 Å). The SEM and EDAX measurements have been performed to probe the surface morphology and layered structure of as-synthesized crystal using the SIGMA VP model High-Resolution Field Emission Scanning Electron Microscope (HR FESEM). The magneto-transport studies were performed via a conventional four-probe method on Quantum Design Physical Property Measurement System (PPMS) equipped with a sample rotator and closed-cycle based cryogen-free system. Magneto-transport measurements were performed at temperature 5K for an applied magnetic field range of ± 10 Tesla.

**Results and Discussion**

The TI and MTI synthesized through a well-optimized solid-state reaction route as described in our previous report [27]. These crystals have been characterized through PXRD (powdered X-ray Diffractometry) and thereby their Rietveld refinement confirms the particular phase of the crystal. The obtained lattice parameters from Rietveld refinement are shown in table 1. It is observed that the lattice parameters increases with increasing tellurium (Te) concentration in $Bi_2Se_3$, which is described in details in our previous report [27]. Using VESTA software, the crystal structures of $Bi_2Te_3$, $Bi_2Se_1Te_2$, $Bi_2Se_2Te_1$, and $Bi_2Se_3$ single crystals (as shown in fig. 1) have been drawn from the Rietveld refinement. The inset of Fig. 1 shows the images of the single crystals of corresponding material. All the four crystals have alternate layers of Bi and Te/Se atoms accordingly. These quintuple layers are held together by weak van der Waals forces. In $Bi_2Se_1Te_2$, as the concentration of Se atoms is increased, they occupy the Te(II) site in $Bi_2Te_3$ crystal. Further increasing the Se concentration, the Se atoms and Te atoms share the Te(I) site in $Bi_2Se_2Te_1$. This confirms the formation of MTI crystals. The Raman spectra have been in direct accordance with the PXRD analysis as there is a shift in Raman mode as we go from $Bi_2Te_3$ to $Bi_2Se_3$, where MTI lies in between. The reason behind these characteristics Raman shift has been explained in our previous report [27].

Fig. 2, shows XRD pattern recorded on crystal flakes of $Bi_2Se_{3-x}Te_x$ (x = 0, 1, 2 & 3). By analyzing thus obtained XRD pattern, we observed that the crystal growth is along the c axis. The peaks have been observed for the (00$3n$) planes only. The particular growth in the (00$l$) plane described the single-crystalline nature of the samples. Comparing different crystals, the peaks have been shifted to the higher angle side in the 2θ axis with an increase in the Se concentration in $Bi_2Se_{3-x}Te_x$. This shifting to lower angle side in XRD pattern is due to doping of Se atom as the Se has a small atomic radius as compared to Te atom. Due to the same reason, there is a huge Raman shift in MTI as compared to TI which is described in our prevoious report [27]. Figure 3 shows SEM images of grown TI and MTI crystals which show that crystals have layered structure which are in direct accordance with the unidirectional growth of crystals. Fig.4 depicts the histogram representation of EDAX measurements for $Bi_2Se_3$, $Bi_2Se_2Te_1$, and $Bi_2Te_3$ TI single crystals. The qualitative weight % of respective atomic constituents (Bi, Se, and Te) were close to the stoichiometric amount of the crystals. It was concluded from the above measurement that all the studied crystals are highly pure and free from impurities.

Fig. 5 represents the temperature-dependent normalized conductivity ($\sigma/\sigma_{230}$) of $Bi_2Se_3$, $Bi_2Se_2Te_1$, $Bi_2Se_1Te_2$, and $Bi_2Te_3$ whereas the inset shows the normalized electrical resistivity ($\rho/\rho_{230}$) for the same down to 10K. The conductivity is maximum for $Bi_2Te_3$ and least for $Bi_2Se_2Te_1$. The obtained values of normalized resistivity and normalized conductivity at temperature 10K for $Bi_2Se_{3-x}Te_x$ (x = 0, 1, 2 & 3) single crystals are shown in table 2. The normalized conductivity at 10K of $Bi_2Te_3$ is approximately four times compared to $Bi_2Se_3$, six times compared to $Bi_2Se_2Te_1$, and three times compared to $Bi_2Se_1Te_2$ TI crystal. While the resistivity-temperature ($\rho$-T) characteristics of all the samples correspond to the metallic behaviour i.e., resistivity decreases with a decrease in temperature. The Residual Resistivity Ratio (RRR) [28,29] value which is taken as the ratio of resistivity at room temperature and 10K ($\rho_{230K}/\rho_{10K}$) is equal to 2.67 for $Bi_2Se_3$, 1.75 for $Bi_2Se_2Te_1$, 3.51 for $Bi_2Se_1Te_2$ and 10.65 for $Bi_2Te_3$ TI crystals. Larger the RRR value, the higher will be the crystalline quality of the sample [30]. The normalized resistivity values at temperature 10K is equal to 0.37 for $Bi_2Se_3$, 0.57 for $Bi_2Se_2Te_1$, 0.29 for $Bi_2Se_1Te_2$ and 0.09 for $Bi_2Te_3$ TI crystals. TI crystals $Bi_2Se_3$ and $Bi_2Te_3$ are having high antisite defects compared to MTI crystals, $Bi_2Se_2Te_1$, and $Bi_2Se_1Te_2$ [31]. Because of this, the TI like $Bi_2Te_3$ is having higher conductivity at low temperature whereas Mixed TI like $Bi_2Se_2Te_1$ is observed to behave as high resistivity material.

Fig. 6 represents the MR % up to ± 10 Tesla magnetic field for $Bi_2Se_{3-x}Te_x$ (x=0-3) TI and MTI crystals at a temperature of 5K. The magneto-resistance (MR) is taken in both directions of the magnetic field and a mean is taken to maintain the symmetry of the final plots. In $Bi_2Se_{3-x}Te_x$ (x = 0 and x = 3) crystals, almost the linear curve dependence of MR on the applied magnetic field has been observed i.e., MR increases with the applied magnetic field in a linear fashion. Further, a clear v-type cusp around the origin is seen in the case of TI like $Bi_2Se_3$ and $Bi_2Te_3$ crystals at lower magnetic fields. This type of behavior is known as the signature of the WAL effect [8,25,32–34]. Interestingly the v- type cusp around the origin in MR% is missing in the case of MTI, i.e., $Bi_2Se_2Te_1$ and $Bi_2Se_1Te_2$. It is clear that not only the magnitude decreases by a huge order but also the shape of the MR% drastically changes from v-type WAL character to U-type, this is a fascinating response of MTIs in comparison to TI. For finding the reason of such ambiguity in MR%, we analyzed the magneto-conductivity data of $Bi_2Se_{3-x}Te_x$ (x = 0-3) in terms of surface conductivity through the HLN hypothesis. The 2-Dimensional (2D) topologically protected non-trivial surface transport properties can be predicted with HLN fitting results [33]. The surface states dominated conductivity of topological insulators is known to follow the HLN equation [35–39] which is given by

$$\Delta\sigma(H) = \sigma(H) - \sigma(0) = -\frac{\propto e^2}{\pi h}\left[\ln(\frac{B_\varphi}{H}) - \Psi\left(\frac{1}{2} + \frac{B_\varphi}{H}\right)\right]$$

Here, $\Delta\sigma(H)$ represents the change in the magneto-conductivity, $\alpha$ is the fitting parameter or prefactor, e denotes the electronic charge, h represents the Planck's constant, $\Psi$ is the digamma function, H is the applied magnetic field. Whereas the characteristic magnetic field is $B_\varphi = \frac{h}{8e\pi l_\varphi^2}$, and $l_\varphi$ is the phase coherence length. The prefactor explains the existence of the WL (Weak Localization) and WAL (Weak Anti Localization) effect while the phase coherence length corresponds to the presence of v-type cusp in the MR curve. The sharpness of cusp decreased with the decrease in the phase coherence length [40]. The WL effect is due to the constructive interference of the electron's wave function between two-time reversal paths, which leads to the localization of electrons. Thereby, this localization affects the ability to transport current in the materials. In the WAL effect, the electron wave function in the two time-reversal path shows destructive interference which leads to the delocalization of electrons in metal [41]. Generally, the $\pi$ Berry phase is associated with the Dirac fermions, which leads to the WAL effect [12]. In the present scenario, the change in MC is negative for the applied magnetic field ($\pm$ 10 Tesla) for all the four Bi-based TI, $Bi_2Se_3$, $Bi_2Se_2Te_1$, $Bi_2Se_1Te_2$, and $Bi_2Te_3$. Thus, it represents the WAL effect is present in these TI crystals which can be regarded as a consequence of $\pi$ Berry phase and the non-trivial topological characteristics of TIs [35].

Figure 7 shows the MC data up to ±10 Tesla with Hikami-Larkin-Nagaoka (HLN) equation fitting up to ±1 Tesla as the HLN only valid in a low magnetic field for all the TIs ($Bi_2Te_3$ and $Bi_2Se_3$) and MTIs ($Bi_2Se_2Te_1$ and $Bi_2Se_1Te_2$) single crystals at 5K. In the HLN, it has been well established that the fitting parameter $\alpha$ signifies the type of localization as it is positive for WL, and negative for WAL [34]. In the ideal system, the value of this prefactor is -0.5 for single coherent topological surface conducting channel that exhibits $\Pi$ Berry phase [33,42–44]. While in the case of multi parallel conducting channels, the pre-factor $\alpha$ value lies between -0.5 to -1.5 [42,43]. However, experimental values of $\alpha$ lied in between –0.3 and –1.1 for the single surface state, two surface states, or intermixing between the surface and bulk states [42,45]. In general, for the WL case, $\alpha$ should be 1, and for the WAL case, $\alpha$ should be -0.5 [24]. In our case of TI and MTI, the magneto-conductivity data is fitted with the HLN equation up to a magnetic field of ±1 Tesla. The prefactor $\alpha$ that comes out through this sophisticated fitting is -0.402 for $Bi_2Se_3$ and -0.336 for $Bi_2Te_3$. The negative value of $\alpha$ describes the WAL effect. The pre-factor $\alpha$ value is nearer to -0.5 which represents a single 2D conduction surface state channel that exists without the backscattering of carriers. At low temperature, the backscattered electrons suffer a destructive interference because of the $\pi$ Berry phase, which increases the conductivity, this is the signature of WAL. Whereas for MTI, at the low temperature, the backscattered electrons obtain constructive interference. This results in a decrease in conductivity logarithmically and corresponds to the signature of WL [46]. The obtained phase coherence length from HLN fitting is 95.11 nm for $Bi_2Se_3$, and 49.53 nm for $Bi_2Te_3$. It is quite clear that in $Bi_2Se_3$, the electrons remain in phase for exactly twice the length as compared to $Bi_2Te_3$.

As the magneto-conductivity data of MTIs does not fit accurately by using the HLN equation in the magnetic field range of ±1 Tesla. Also, the fitted curve has a negative value of $R^2$ (goodness of fit) which suggests that the HLN is not applicable for MTIs. Therefore, it is not accurate to say anything about WL and WAL effect for MTIs based on HLN fitting. To study this anomalous behaviour in MTIs, we fit the magneto-conductivity data with the $\beta H^2$ term. This particular term signifies the bulk contribution as a whole from the system [32,47,48]. Here the coefficient $\beta$ represents the contribution of both classical cyclotronic and quantum scattering term [49–51]. The MC data of MTIs is well fitted by the $\beta H^2$ term as shown in figure 8, which suggests the dominance of bulk conduction. The obtained values of $\beta$ by fitting are $-1.091 \times 10^{-4}$ and $-9.762 \times 10^{-5}$ for $Bi_2Se_2Te_1$ and $Bi_2Se_1Te_2$ respectively. The magneto-resistance of MTIs shows the quadratic dependence on the applied magnetic field and thereby magneto-conductivity also has a similar trend too. The absence of sharp v-type cusp in the magneto-conductivity of MTIs also confirms the suppression of surface contribution to conductivity that represents a crossover from surface charge carrier driven conductivity in pristine crystals to bulk charge carrier conductivity in the case of MTI.

**Conclusion**

We successfully synthesized the $Bi_2Se_{3-x}Te_x$ (x = 0-3) TI crystals. XRD measurement on crystal flakes confirms the (00*3n*) unidirectional growth of all the four TI crystal samples. The surface morphology has been analyzed with SEM and confirms the layered structure of all the crystals. In the RT measurements, pure TI crystals show high conductivity at low temperatures whereas MTI crystals have unusual high resistance. A negative MC is observed in all samples, and we successfully fitted the MC with the HLN equation. In both $Bi_2Se_3$ and $Bi_2Te_3$ TI crystals, there exists a WAL phenomenon. A clear v-type cusp with linear MR % behaviour is observed in these pure TI samples. MTI crystals like $Bi_2Se_2Te_1$ and $Bi_2Se_1Te_2$ showed parabolic MC characteristics and HLN is not applicable here. So the MC data have been fitted with $\beta H^2$ which represent dominance of bulk charge carriers. Hence we concluded that in pristine crystals the surface contribution is dominant whereas in MTIs the bulk contribution overtakes the conducting phenomena.

**Acknowledgment**

The authors would like to thank Director CSIR-NPL for his keen interest and encouragement. The authors are also grateful to Poonam Rani, Pankaj Maheshwari, Prince Sharma, Nav Jyoti (from JMI, New Delhi) for their help in the writing the manuscript, preparation of figures, and helpful interactions and other additional support. Deepak Sharma, Yogesh Kumar would like to thanks CSIR for the research fellowship and AcSIR-Ghaziabad for Ph.D. registration.

**Table 1:** Lattice parameters of $Bi_2Se_3$, $Bi_2Se_2Te_1$, $Bi_2Se_1Te_2$ and $Bi_2Te_3$ single crystals obtained from Reitveld refinement.

|  | a (Å) | b (Å) | c (Å) | α, β | γ |
|---|---|---|---|---|---|
| $Bi_2Se_3$ | 4.156(3) | 4.156(3) | 28.736(8) | $90^0$ | $120^0$ |
| $Bi_2Se_2Te_1$ | 4.219(3) | 4.219(3) | 29.511(5) | $90^0$ | $120^0$ |
| $Bi_2Se_1Te_2$ | 4.281(4) | 4.281(4) | 29.922(5) | $90^0$ | $120^0$ |
| $Bi_2Te_3$ | 4.386(6) | 4.386(6) | 30.499(1) | $90^0$ | $120^0$ |

**Table 2:** Normalized resistivity ($\rho_{10K}/\rho_{230K}$) and normalized conductivity ($\sigma_{10K}/\sigma_{230K}$) at temperature 10K of $Bi_2Se_3$, $Bi_2Se_2Te_1$, $Bi_2Se_1Te_2$ and $Bi_2Te_3$ single crystals.

|  | ($\rho_{10K}/\rho_{230K}$) | ($\sigma_{10K}/\sigma_{230K}$) |
|---|---|---|
| $Bi_2Se_3$ | 0.37 | 2.67 |
| $Bi_2Se_2Te_1$ | 0.57 | 1.75 |
| $Bi_2Se_1Te_2$ | 0.29 | 3.51 |
| $Bi_2Te_3$ | 0.09 | 10.65 |

**Figure captions**

Fig.1: Unit Cell of single crystals (a) $Bi_2Se_3$, (b) $Bi_2Se_2Te_1$, (c) $Bi_2Se_1Te_2$ and (d) $Bi_2Te_3$. The inset of each structure shows the image of the as-synthesized single crystal.

Fig.2: X-ray diffraction pattern for $Bi_2Se_{3-x}Te_x$ (x= 0, 1, 2 & 3) single crystals showin the (00l) planes.

Fig.3: Scanning Electron Microscopic images of (a) $Bi_2Se_3$, (b) $Bi_2Se_2Te_1$, (c) $Bi_2Se_1Te_2$, and (d) $Bi_2Te_3$.

Fig.4: Elemental analysis of pristine and MTI crystals by Energy-dispersive X-ray spectroscopy.

Fig.5: Temperature-dependent normalized magneto-conductivity ($\sigma/\sigma_{230}$) of $Bi_2Se_3$, $Bi_2Se_2Te_1$, $Bi_2Se_1Te_2$, and $Bi_2Te_3$, inset shows the normalized electrical resistivity ($\rho/\rho_{230}$) of the same.

Fig.6: MR (%) as a function of magnetic field for (a) $Bi_2Se_3$ (b) $Bi_2Se_2Te_1$ (c) $Bi_2Se_1Te_2$ and (d) $Bi_2Te_3$ at 5K.

Fig.7: HLN fitted plot of magneto-conductivity measurements in the field range of ± 10 Tesla for (a) $Bi_2Se_3$ (b) $Bi_2Se_2Te_1$ (c) $Bi_2Se_1Te_2$ and (d) $Bi_2Te_3$ crystals at 5K temperature.

Fig.8: $\beta H^2$ fitted plot of magneto-conductivity measurement for $Bi_2Se_2Te_1$ and $Bi_2Se_1Te_2$ at temperature 5K and field range of ± 10 Tesla.

**References**

1. M. He, H. Sun, and Q. L. He, "Topological insulator: Spintronics and quantum computations," Front. Phys. **14**(4), 43401 (2019).
2. S. Ornes, "Topological insulators promise computing advances, insights into matter itself," Proc. Natl. Acad. Sci. U. S. A. **113**(37), 10223–10224 (2016).
3. Y. Fan and K. L. Wang, "Spintronics based on topological insulators," Spin **6**(2), 1640001 (2016).
4. J. E. Moore, "The birth of topological insulators," Nature **464**(7286), 194–198 (2010).
5. W. Tian, W. Yu, J. Shi, and Y. Wang, "The property, preparation and application of topological insulators: A review," Materials (Basel). **10**(7), 814 (2017).
6. D. O. Scanlon, P. D. C. King, R. P. Singh, A. De La Torre, S. M. K. Walker, G. Balakrishnan, F. Baumberger, and C. R. A. Catlow, "Controlling bulk conductivity in topological insulators: Key role of anti-site defects," Adv. Mater. **24**(16), 2154–2158 (2012).
7. R. Sultana, D. Sharma, R. S. Meena, and V. P. S. Awana, "Signatures of Quantum Transport Steps in Bi2Se3 Single Crystal," J. Supercond. Nov. Magn. **32**(6), 1497–1499 (2019).
8. D. Sharma, P. Rani, P. K. Maheshwari, V. Nagpal, R. S. Meena, S. S. Islam, S. Patnaik, and V. P. S. Awana, "Hikami-Larkin-Nagaoka (HLN) fitting of magneto transport of Bi2Se3 single crystal in different magnetic field ranges," in *3Rd International Conference on Condensed Matter and Applied Physics (Icc-2019)* (2020), **2220**, p. 110028.
9. G. Tkachov, *Topological Insulators: The Physics of Spin Helicity in Quantum Transport* (Jenny Stanford Publishing, 2015).
10. D. Kong and Y. Cui, "Opportunities in chemistry and materials science for topological insulators and their nanostructures," Nat. Chem. **3**(11), 845–849 (2011).
11. M. Z. Hasan and C. L. Kane, "Colloquium: Topological insulators," Rev. Mod. Phys. **82**(4), 3045–3067 (2010).
12. Y. Ando, P. Society, R. April, and Y. Ando, "Topological Insulator Materials," J. Phys. Soc. Japan **82**(10), 1–32 (2013).
13. H. Cao, J. Tian, I. Miotkowski, T. Shen, J. Hu, S. Qiao, and Y. P. Chen, "Quantized hall effect and shubnikov-de haas oscillations in highly doped Bi 2Se 3: Evidence for layered transport of bulk carriers," Phys. Rev. Lett. **108**(21), 216803 (2012).
14. W. Wang, Y. Du, G. Xu, X. Zhang, E. Liu, Z. Liu, Y. Shi, J. Chen, G. Wu, and X. X. Zhang, "Large linear magnetoresistance and shubnikov-de hass oscillations in single crystals of YPdBi heusler topological insulators," Sci. Rep. **3**(1), 2181 (2013).
15. H. Liu, S. Liu, Y. Yi, H. He, and J. Wang, "Shubnikov-de Haas oscillations in n and p type Bi2Se3 flakes," 2D Mater. **2**(4), 045002 (2015).
16. H. Wang, Y. Liu, Y. Liu, C. Xi, J. Wang, J. Liu, Y. Wang, L. Li, S. P. Lau, M. Tian, J. Yan, D. Mandrus, J. Y. Dai, H. Liu, X. Xie, and J. Wang, "Log-periodic quantum magneto-oscillations and discrete-scale invariance in topological material HfTe5," Natl. Sci. Rev. **6**(5), 914–920 (2019).
17. W. J. Wang, K. H. Gao, Z. Q. Li, T. Lin, J. Li, C. Yu, and Z. H. Feng, "Classical linear magnetoresistance in epitaxial graphene on SiC," Appl. Phys. Lett. **105**(18), 182102 (2014).
18. S. Singh, R. K. Gopal, J. Sarkar, A. Pandey, B. G. Patel, and C. Mitra, "Linear magnetoresistance and surface to bulk coupling in topological insulator thin films," J. Phys. Condens. Matter **29**(50), 505601 (2017).
19. Y. Shiomi and E. Saitoh, "Linear magnetoresistance in a topological insulator Ru2Sn3," AIP Adv.
7 | P a g e

**7**(3), 035011 (2017).
20. M. Li, Z. Wang, L. Yang, X. P. A. Gao, and Z. Zhang, "From linear magnetoresistance to parabolic magnetoresistance in Cu and Cr-doped topological insulator Bi2Se3 films," J. Phys. Chem. Solids **128**, 331–336 (2019).
21. M. Petrushevsky, E. Lahoud, A. Ron, E. Maniv, I. Diamant, I. Neder, S. Wiedmann, V. K. Guduru, F. Chiappini, U. Zeitler, J. C. Maan, K. Chashka, A. Kanigel, and Y. Dagan, "Probing the surface states in Bi 2Se 3 using the Shubnikov-de Haas effect," Phys. Rev. B - Condens. Matter Mater. Phys. **86**(4), 045131 (2012).
22. H. Zhang, C.-X. Liu, X.-L. Qi, X. Dai, Z. Fang, and S.-C. Zhang, "Topological Insulators in Bi2Se3, Bi2Te3 and Sb2Te3 with single Dirac cone on the surface," Nat. Phys. **5**, 438–442 (2009).
23. K. Park, Y. Nomura, R. Arita, A. Llobet, and D. Louca, "Local strain and anharmonicity in the bonding of Bi 2 Se 3 - X Te x topological insulators," Phys. Rev. B - Condens. Matter Mater. Phys. **88**(22), 224108 (2013).
24. L. Bao, L. He, N. Meyer, X. Kou, P. Zhang, Z. G. Chen, A. V. Fedorov, J. Zou, T. M. Riedemann, T. A. Lograsso, K. L. Wang, G. Tuttle, and F. Xiu, "Weak anti-localization and quantum oscillations of surface states in topological insulator Bi 2Se 2Te," Sci. Rep. **2**(1), 726 (2012).
25. E. P. Amaladass, T. R. Devidas, S. Shilpam, C. S. Sundar, M. Awadhesh, and A. Bharathi, "Magneto-transport behaviour of Bi 2 Se 3− x Te x : role of disorder," J. Phys. Condens. Matter **28**(7), 75003 (2016).
26. B. A. Assaf, T. Cardinal, P. Wei, F. Katmis, J. S. Moodera, and D. Heiman, "Linear magnetoresistance in topological insulator thin films: Quantum phase coherence effects at high temperatures," Appl. Phys. Lett. **102**(1), 012102 (2013).
27. D. Sharma, M. M. Sharma, R. S. Meena, and V. P. S. Awana, "Raman Spectroscopy of Bi 2 Se 3-x Te x ( x = 0-3 ) Topological Insulator Crystals," Physica B-Condensed matter physics (August 2020).
28. R. Sultana, G. Gurjar, S. Patnaik, and V. P. S. Awana, "Crystal growth and characterization of bulk Sb2Te3 topological insulator," Mater. Res. Express **5**(4), 046107 (2018).
29. R. Sultana, G. Gurjar, B. Gahtori, S. Patnaik, and V. P. S. Awana, "Flux free single crystal growth and detailed physical property characterization of Bi1-xSbx (x = 0.05, 0.1 and 0.15) topological insulator," Mater. Res. Express **6**(10), 106102 (2019).
30. K. Shrestha, M. Chou, D. Graf, H. D. Yang, B. Lorenz, and C. W. Chu, "Extremely large nonsaturating magnetoresistance and ultrahigh mobility due to topological surface states in the metallic Bi2Te3 topological insulator," Phys. Rev. B **95**(19), 195113 (2017).
31. R. J. Cava, H. Ji, M. K. Fuccillo, Q. D. Gibson, and Y. S. Hor, "Crystal structure and chemistry of topological insulators," J. Mater. Chem. C **1**(19), 3176–3189 (2013).
32. C. Shekhar, C. E. Violbarbosa, B. Yan, S. Ouardi, W. Schnelle, G. H. Fecher, and C. Felser, "Evidence of surface transport and weak antilocalization in a single crystal of the Bi2 Te2Se topological insulator," Phys. Rev. B - Condens. Matter Mater. Phys. **90**(16), 165140 (2014).
33. G. M. Stephen, O. A. Vail, J. Lu, W. A. Beck, P. J. Taylor, and A. L. Friedman, "Weak Antilocalization and Anisotropic Magnetoresistance as a Probe of Surface States in Topological Bi2TexSe3−x Thin Films," Sci. Rep. **10**(1), 4845 (2020).
34. H. Z. Lu, J. Shi, and S. Q. Shen, "Competition between weak localization and antilocalization in topological surface states," Phys. Rev. Lett. **107**(7), 076801 (2011).
35. M. Liu, J. Zhang, C. Z. Chang, Z. Zhang, X. Feng, K. Li, K. He, L. L. Wang, X. Chen, X. Dai, Z.

Fang, Q. K. Xue, X. Ma, and Y. Wang, "Crossover between weak antilocalization and weak localization in a magnetically doped topological insulator," Phys. Rev. Lett. **108**(3), 036805 (2012).
36. F. E. Meijer, A. F. Morpurgo, T. M. Klapwijk, and J. Nitta, "Universal spin-induced time reversal symmetry breaking in two-dimensional electron gases with rashba spin-orbit interaction," Phys. Rev. Lett. **94**(18), 186805 (2005).
37. S. Hikami, A. I. Larkin, and Y. Nagaoka, "Spin-Orbit Interaction and Magnetoresistance in the Two Dimensional Random System," Prog. Theor. Phys. **63**(2), 707–710 (1980).
38. J. Chen, X. Y. He, K. H. Wu, Z. Q. Ji, L. Lu, J. R. Shi, J. H. Smet, and Y. Q. Li, "Tunable surface conductivity in Bi2Se3 revealed in diffusive electron transport," Phys. Rev. B - Condens. Matter Mater. Phys. **83**(24), 241304 (2011).
39. W. E. Liu, E. M. Hankiewicz, and D. Culcer, "Weak localization and antilocalization in topological materials with impurity spin-orbit interactions," Materials (Basel). **10**(7), 807 (2017).
40. H. T. He, G. Wang, T. Zhang, I. K. Sou, G. K. L. Wong, J. N. Wang, H. Z. Lu, S. Q. Shen, and F. C. Zhang, "Impurity effect on weak antilocalization in the topological insulator Bi2Te3," Phys. Rev. Lett. **106**(16), 166805 (2011).
41. P. A. Lee and T. V. Ramakrishnan, "Disordered electronic systems," Rev. Mod. Phys. **57**(2), 287–337 (1985).
42. H. Z. Lu and S. Q. Shen, "Weak localization of bulk channels in topological insulator thin films," Phys. Rev. B - Condens. Matter Mater. Phys. **84**(12), 125138 (2011).
43. B. Irfan, "Surface characterization and magneto-transport study on Bi 2Te 2Se topological insulator thin film," Appl. Phys. A Mater. Sci. Process. **126**(3), 191 (2020).
44. H. Steinberg, J. B. Laloë, V. Fatemi, J. S. Moodera, and P. Jarillo-Herrero, "Electrically tunable surface-to-bulk coherent coupling in topological insulator thin films," Phys. Rev. B - Condens. Matter Mater. Phys. **84**(23), 233101 (2011).
45. H.-Z. Lu and S.-Q. Shen, "Weak localization and weak anti-localization in topological insulators," in *Spintronics VII*, H.-J. Drouhin, J.-E. Wegrowe, and M. Razeghi, eds. (2014), **9167**, p. 91672E.
46. H. Z. Lu and S. Q. Shen, "Finite-temperature conductivity and magnetoconductivity of topological insulators," Phys. Rev. Lett. **112**(14), 146601 (2014).
47. W. H. Shon and J. S. Rhyee, "Fermi level tuning and weak localization/weak antilocalization competition of bulk single crystalline Bi2-xSbxSe2Te compounds," J. Phys. Condens. Matter **27**(2), 025502 (2015).
48. B. Bhattacharyya, A. Sharma, V. P. S. Awana, T. D. Senguttuvan, and S. Husale, "FIB synthesis of Bi2Se3 1D nanowires demonstrating the co-existence of Shubnikov-de Haas oscillations and linear magnetoresistance," J. Phys. Condens. Matter **29**(7), 07LT01 (2017).
49. E. P. Amaladass, T. R. Devidas, S. Sharma, and A. Mani, "Quantum coherence phenomenon in disordered Bi2SeTe2 topological single crystal: Effect of annealing," J. Phys. Condens. Matter **29**(17), 175602 (2017).
50. R. K. Gopal, S. Singh, A. Mandal, J. Sarkar, and C. Mitra, "Topological delocalization and tuning of surface channel separation in Bi2 Se2 Te Topological Insulator Thin films," Sci. Rep. **7**(1), 4924 (2017).
51. R. K. Gopal, S. Singh, J. Sarkar, and C. Mitra, "Tuning chemical potential in the Dirac cone by compositional engineering," AIP Adv. **7**(10), 105112 (2017).

*Fig. 1*

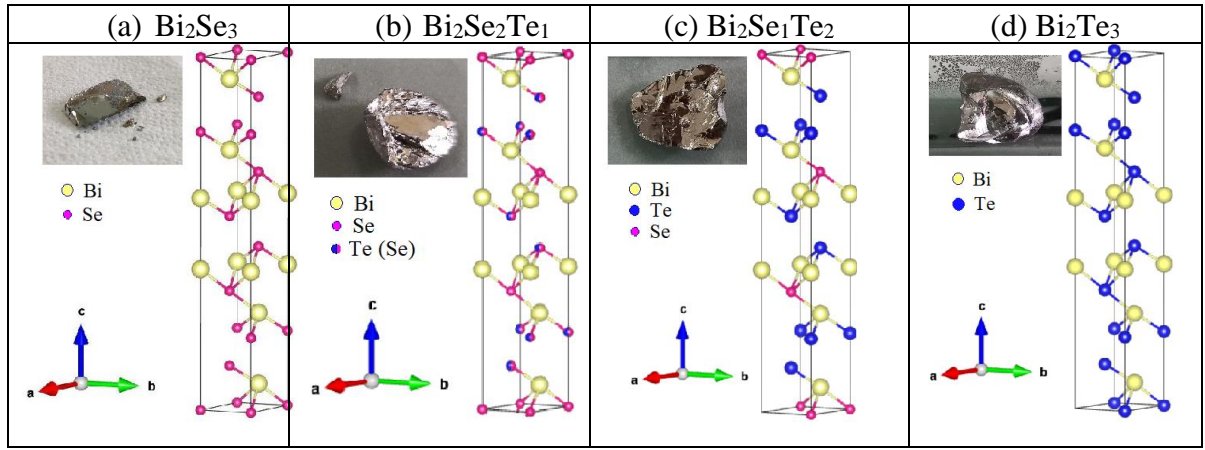

*Fig. 2*

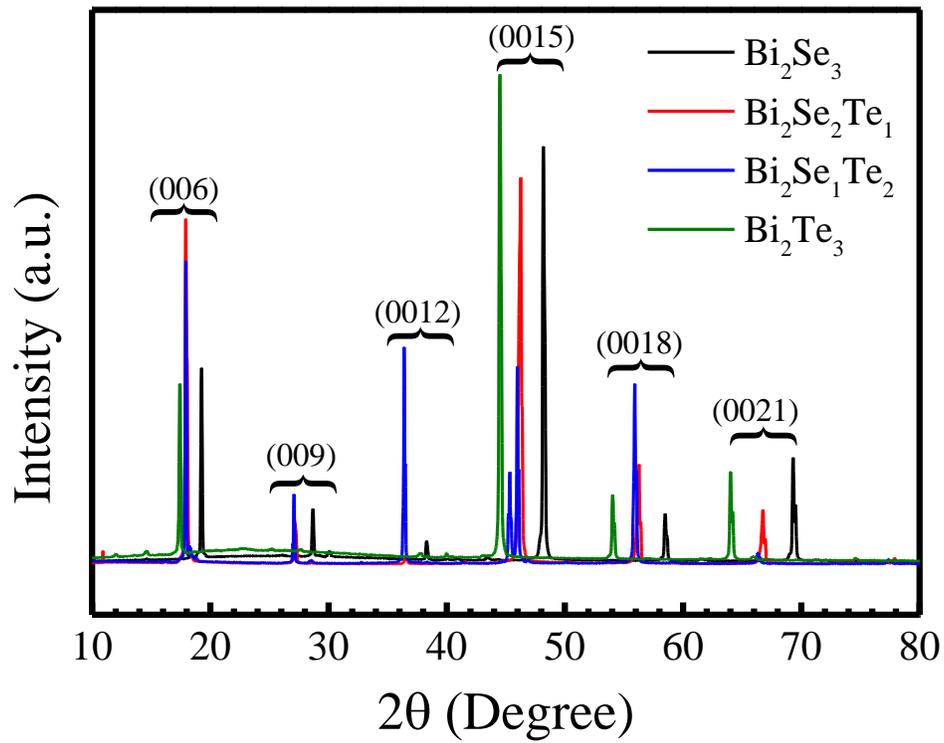

*Fig. 3*

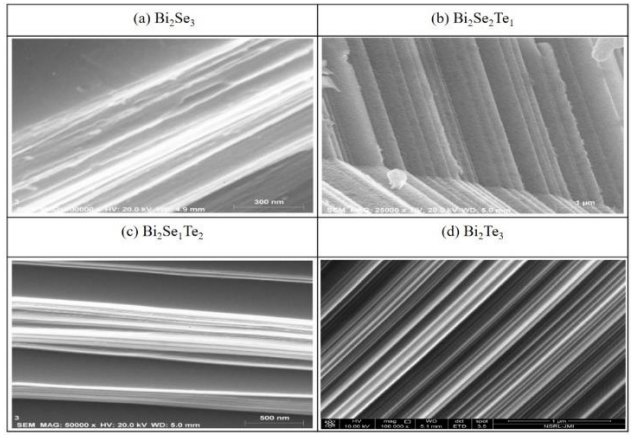

*Fig. 4*

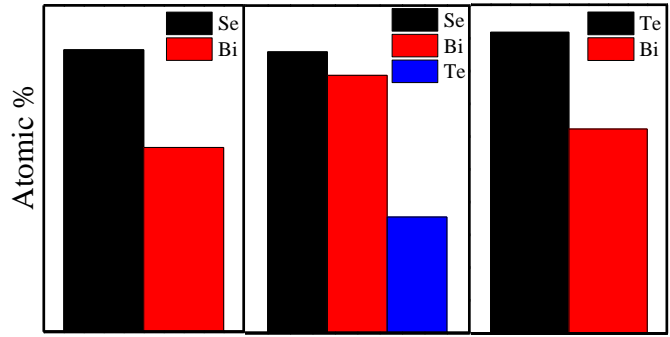

*Fig. 5*

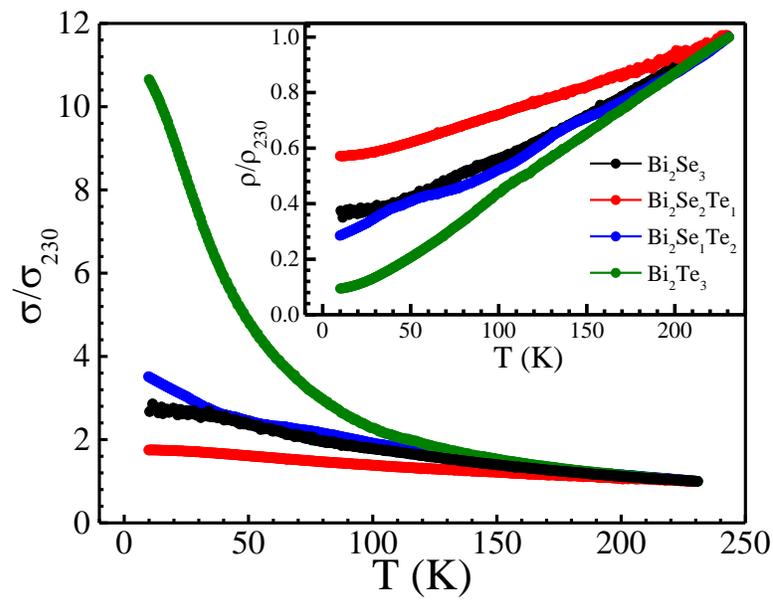

*Fig. 6*

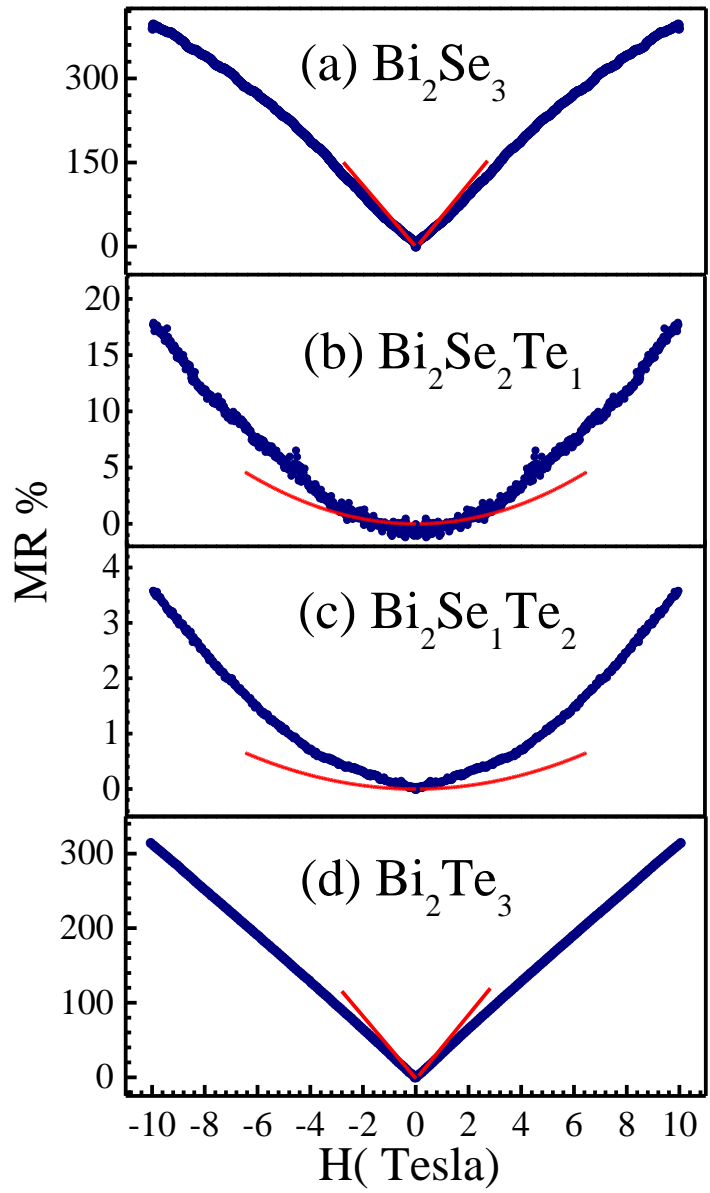

*Fig. 7*

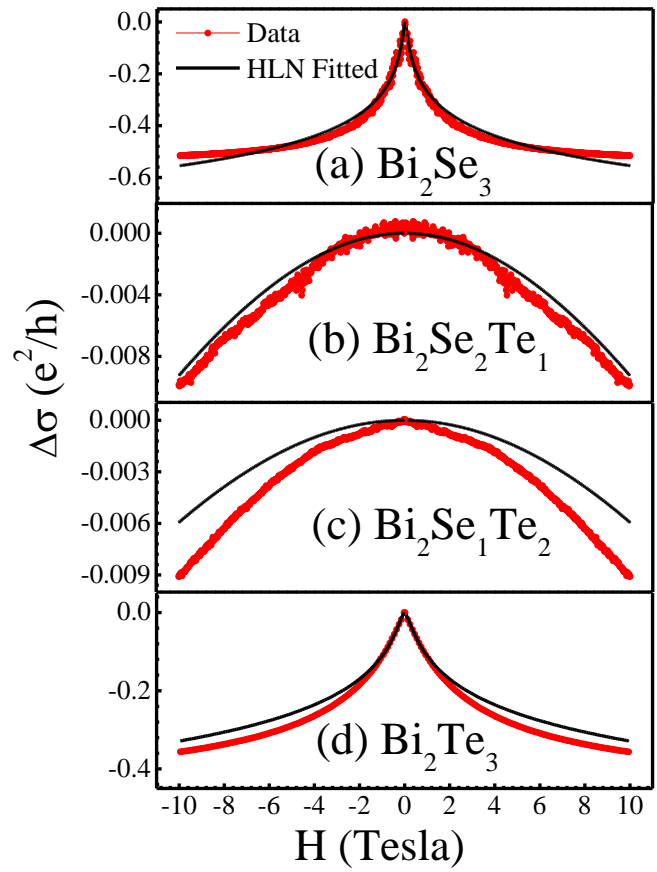

*Fig. 8*

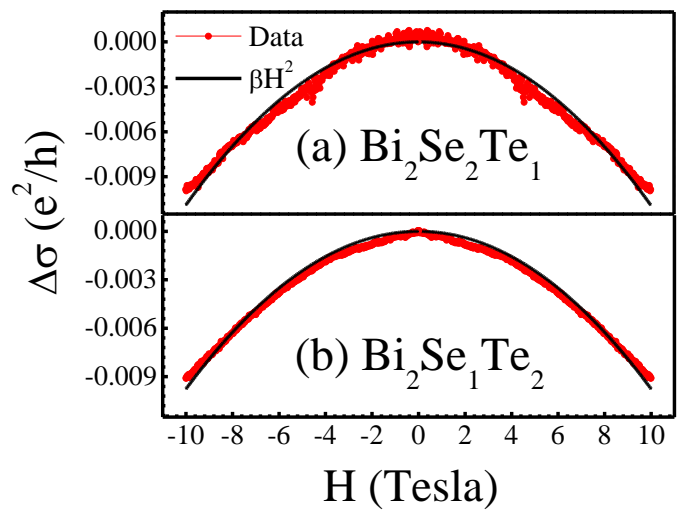